\documentclass[prd, twocolumn, nofootinbib, floatfix]{revtex4-1}

\usepackage{amsmath}
\usepackage{graphicx}
\usepackage{dcolumn}
\usepackage{bm}
\usepackage{epsfig}
\usepackage{amssymb,latexsym,mathrsfs}
\usepackage{graphicx}
\usepackage{color}
\usepackage{hyperref}

\hypersetup{
    colorlinks=true,
    linkcolor=red,
    citecolor=blue,
} 

\newcommand{\be}{\begin{equation}}
\newcommand{\ee}{\end{equation}}
\newcommand{\bs}{\begin{split}} 
\newcommand{\bea}{\begin{eqnarray}}
\newcommand{\eea}{\end{eqnarray}}

\newcommand{\om}{\Omega_m}

\newcommand{\Oe}{\Omega_{e}}
 
\newcommand{\neff}{N_{\rm eff}} 
\newcommand{\hfid}{H^2_{\rm fid}}

\begin{document}

\title{Model Independent Early Expansion History and Dark Energy} 
\author{Johan Samsing$^{1,2}$, Eric V.\ Linder$^{1,3}$, Tristan L.\ 
Smith$^1$} 
\affiliation{$^1$Berkeley Center for Cosmological Physics \& Berkeley Lab, 
University of California, Berkeley, CA 94720, USA\\ 
$^2$Dark Cosmology Centre, Niels Bohr Institute, University of Copenhagen, 
Juliane Maries Vej 30, 2100 Copenhagen, Denmark\\ 
$^3$Institute for the Early Universe WCU, Ewha Womans University, 
Seoul, Korea}

\begin{abstract}
We examine model independent constraints on the high redshift and 
prerecombination expansion history from cosmic microwave background 
observations, 
using a combination of principal component analysis and other techniques.  
This can be translated to model independent limits on early dark energy and 
the number of relativistic species $\neff$.  Models such as scaling 
(Doran-Robbers), dark radiation ($\Delta\neff$), and barotropic aether 
fall into distinct regions of eigenspace and can be easily distinguished 
from each other.  
Incoming CMB data will map the expansion history from $z=0$--$10^5$, 
achieving subpercent precision around recombination, and 
enable determination of the amount of early dark 
energy and valuable guidance to its nature.  
\end{abstract}

\date{\today} 

\maketitle

%%%%%%%%%%%%%%%%%%%%%%%%%%%%%%%%%%%%%%%%%%%%%%%%%%%%%%%%%%%%%%%%%%%%%%%%
\section{Introduction} 

The expansion history of the universe is a fundamental property of 
cosmology, reflecting the energy density constituents and their evolution. 
Yet remarkably little is known in detail about it, other than in a 
coarse grained average.  For redshifts between 3000 and $10^9$, the universe 
was mostly radiation dominated, for redshifts between 3000 and $\sim1$ it 
was mostly matter dominated, but excursions are possible -- in the effective 
number of relativistic species $\neff$ say, or even temporary breakdown of 
such domination -- and the level of subdominant components is not well 
constrained.  
Only around the epoch of primordial nucleosynthesis and of recombination is 
the expansion rate (Hubble parameter) better constrained, but even there at 
the $\sim5\%$ level averaged over the epoch \cite{kaplinghat,zahn}. 

Given the importance of the expansion history, and the improvement in 
cosmic microwave background (CMB) data, we investigate what constraints 
can be placed on it in a model independent way, i.e.\ other than fitting 
for a deviation of a particular functional form such as extra $\neff$ or 
a specific dark energy model.  
This would fill in a vast range of cosmic expansion where almost no precise 
constraints have been placed.  That is, an error band for the Hubble parameter 
$H(z)$ at $z>1000$ should be a staple of cosmology textbooks, and yet 
does not exist. 

The early expansion history has an important bearing on understanding 
the nature of dark energy as well, the question of {\it persistence\/} 
of dark energy.  For a cosmological constant $\Lambda$, the dark energy 
density contributed at recombination is $\Omega_\Lambda\approx 10^{-9}$, 
while the current upper limit from data is above $10^{-2}$.  This gives 
substantial unexplored territory.  Moreover, the current constraints use 
a specific functional form for the dark energy evolution (usually the 
Doran-Robbers form \cite{dorrob}), but other models could lead to 
significantly different limits \cite{rdof}.  Thus, model independent limits 
on early 
dark energy are needed.  Physics origins for early dark energy can be 
quite diverse, e.g.\ from dilaton models (as in some string theories) to 
k-essence (noncanonical kinetic field theories) to dark radiation (as in 
some higher dimension theories) \cite{cope}.  Establishing whether CMB 
observations 
could distinguish these classes is another important question. 

Improvement of CMB data recently by higher resolution observations 
extending the temperature power spectrum to multipoles $\ell\approx3000$ 
by the Atacama Cosmology Telescope (ACT \cite{act}) and South Pole 
Telescope (SPT \cite{spt}) gives valuable leverage since higher multipoles 
are sensitive to modes crossing the cosmological horizon at earlier times. 
This advance was used in \cite{linsmith} to rule out in a model independent 
manner the presence of any epoch of cosmic acceleration between $z\approx2$ 
and $10^5$ (supplementing the limits from growth of structure 
post-recombination in \cite{unique}).  Upcoming Planck and ground based 
polarization experiment data will also map out the polarization power 
spectra, giving additional constraints. 

To carry out a model independent analysis of the early expansion history, 
we use a combination of redshift binning and principal component analysis. 
In Sec.~\ref{sec:bins} we lay out the methodology for describing arbitrary 
$H(z)$.  Analyzing the results in Sec.~\ref{sec:pca}, we identify the 
redshifts ranges where the CMB observations are most sensitive to expansion 
variations.  We project three classes of models representing different 
physical origins onto the eigenmodes to explore the discriminating power of 
the data in Sec.~\ref{sec:edemodels}.  In Sec.~\ref{sec:concl} we discuss 
the results and future prospects.

%%%%%%%%%%%%%%%%%%%%%%%%%%%%%%%%%%%%%%%%%%%%%%%%%%%%%%%%% 
\section{Expansion History} \label{sec:bins} 

The expansion Friedmann equation directly relates the expansion rate 
of the universe, or Hubble parameter, to the energy density constituents, 
\be 
H^2(a)=\frac{8\pi G}{3}\sum \rho_i(a) \ , 
\ee 
where we neglect curvature (from a theoretical prior for flatness 
and because we mostly treat high redshift where it would be negligible). 
At high redshift the canonical expectation is that the universe is matter 
or radiation dominated, so we write 
\bea 
H^2(a)&=&\frac{8\pi G}{3}\,[\rho_m(a)+\rho_r(a)+\rho_\Lambda]+\delta H^2(a)\\ 
&=&\hfid+\delta H^2(a)\\ 
&=&\hfid\,[1+\delta(a)] \ . 
\eea 

Deviations $\delta(a)$ to the fiducial expansion rate can also be 
interpreted as an effective dark energy density differing from that of the 
cosmological constant, with 
\be 
\rho_{de}(a)=\rho_\Lambda+\frac{3\hfid}{8\pi G}\,\delta \ . 
\ee 
We can write the dark energy density evolution as 
\bea 
\rho_{de}(a)&=&\rho_{de,0}\,f(a)=\rho_\Lambda\,f(a)\\ 
f(a)&=&1+\left(1+\frac{\rho_{bg}}{\rho_\Lambda}\right)\,\delta \ , 
\eea 
where $\rho_{bg}$ is the background energy density excepting dark energy, 
i.e.\ usually the dominant component, 
matter or radiation.  More simply, the fraction of critical density 
contributed by the effective dark energy is 
\be 
\Omega_{de}(a)=\frac{8\pi G\rho_\Lambda}{3\hfid\,(1+\delta)}+ 
\frac{\delta}{1+\delta} = \frac{\Omega_\Lambda(a)+\delta}{1+\delta} \ . 
\ee 
We can readily see that at high redshift we obtain a fractional early 
dark energy 
density contribution of approximately $\delta(a)$, for 
$\Omega_\Lambda(a)\ll \delta\ll 1$.  
During epochs when $\delta$ is constant, this is a constant fractional 
contribution.

Our goal is to analyze constraints on the variations $\delta(a)$ from 
the canonical model with $\delta=0$.  We begin by writing $\delta(a)$ 
as a linear combination in an orthogonal bin basis, 
\be 
\delta(a)=\sum \beta_i b_i(a) \ , \label{eq:binh} 
\ee 
where $b_i(a)$ is a tophat of amplitude 1 over a given range of scale 
factor $a$, and 0 otherwise.  That is, the Hubble parameter deviations 
$\delta(a)$ are given as a linear combination of piecewise constant values. 
We can then constrain $H(a)$ in bins of $a$, a model independent 
description.  The bin basis is also the standard first step in principal 
component analysis (see, e.g., \cite{hutstar}), as we will pursue in the 
next section.  We choose $N$ bins per decade of scale factor over the 
range of $\log a=[-5,0]$, beginning with $N=20$. 

Since we are interested in the expansion history we deal directly with 
the Hubble parameter (or effective dark energy density).  Treating bins 
of the dark energy equation of state, or pressure to density, ratio $w(z)$ 
would have some drawbacks here.  Most severe is that to obtain $H(z)$ one 
must integrate 
$w(z')$ over all redshifts from zero to $z$.  This makes it difficult to 
explore the early expansion history in a model independent manner.  Moreover, 
the instantaneous $w(z)$ is not fully informative: during matter domination, 
for example, any level of dark energy density from $\Omega_{de}=10^{-9}$ 
to $10^{-2}$ or whatever that scales as the matter has $w=0$.  Thus we 
aim to derive constraints directly on variations in $H(z)$, and consider the 
interpretation of these as a further step. 

The expansion history directly feeds into the CMB power spectra, through 
changing the distance scales, e.g.\ of the sound horizon or damping scale, 
and the relation of multipole $\ell$ (or angular scale $\theta$) to wavenumber 
$k=\ell/\eta(z)$, where $\eta$ is the conformal distance.  It also affects 
the decoupling of photons from baryons and the growth of perturbations in 
both.  

The treatment of perturbations requires some attention.  The description of 
the cosmic expansion gives the evolution of the homogeneous background, but 
consistency of the field equations requires consideration of perturbations 
in all components of energy density.  Unless the deviation in $H(z)$ is 
interpreted purely in terms of a cosmological constant (which indeed is 
purely homogeneous), spatial perturbations have to be accounted for, at least 
formally and generally in actual practice.  The perturbation evolution 
equations for the additional energy density involve the quantities $w(z)$, 
$w'(z)=dw/d\ln a$, the initial conditions on the density perturbation, and 
the sound speed of the effective fluid $c_s(z)$.  (One could also add a 
viscous sound speed or anisotropic stresses, see \cite{hu1998}.) 

The first three of these are fairly straightforward.  For any deviation 
$\delta(a)$ one can define an effective equation of state 
\bea 
w&=&-1-\frac{1}{3}\frac{d\ln\rho_{de}}{d\ln a}\\ 
&=&-1-\frac{a}{3[1+\delta(1+\rho_{bg}/\rho_\Lambda)]} 
\left[\left(1+\frac{\rho_{bg}}{\rho_\Lambda}\right)\frac{d\delta}{da} + \frac{\delta}{\rho_\Lambda}\frac{d\rho_{bg}}{da}\right]\\ 
&\approx& w_{bg}-\frac{1}{3}\frac{d\ln\delta}{d\ln a} \ , 
\eea 
where the last line holds when $\rho_\Lambda$ gives negligible contribution 
to the effective dark energy density.  One can take a further derivative 
to obtain $w'$.  
Initial conditions 
are usually taken as adiabatic and stresses are taken to vanish.  However, 
one does have to specify the sound speed.  If one interprets the extra 
energy density as arising from quintessence, i.e.\ a minimally coupled, 
canonical scalar field, then $c_s=1$.  In general, the necessary inclusion 
of perturbations in whatever is the origin of the deviations in the expansion 
history prevents a purely model independent treatment -- one has to make 
some assumptions about the physics.  
Here we fix $c_s$ to that for the 
particular cases we consider, but in future work we will fit for it. 

We modified CAMB \cite{camb} to allow a general $H(z)$, with the $w(z)$ 
that goes along with this. 
We then solve the coupled background evolution, and 
photon, matter, and effective dark energy perturbations equations to obtain 
the CMB power spectra. 
For evaluating binned $H(z)$ models, using the orthogonal bin basis 
introduced in Eq.~(\ref{eq:binh}), we slightly smooth the bin edges, using 
a Gaussian smoothing of width $0.075$ times the bin width, to prevent 
infinite derivatives.  We extensively test convergence and numerical 
stability of the results (also see \cite{linsmith} where this procedure 
was found to be robust).  

\begin{figure}[htbp!]
\includegraphics[width=\columnwidth]{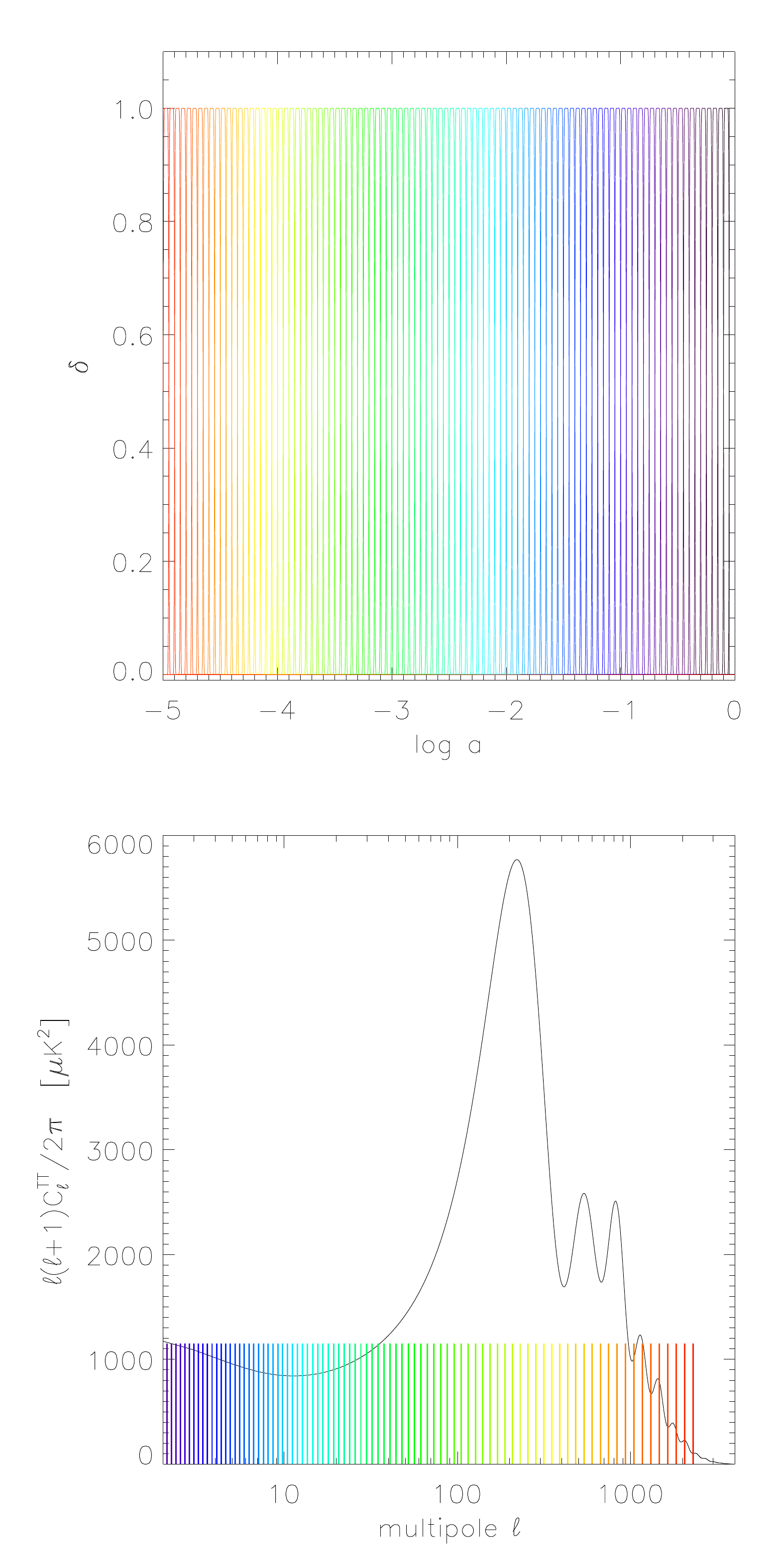}
\caption{[Top] The bin basis for expansion variations $\delta(a_i)$ 
is plotted vs $\log a$.  [Bottom] The scale factors $a_i$ of the center 
of each bin in the top panel (with matching colors) are approximately 
mapped into multipoles by $\ell\approx\eta_0/\eta(a_i)$, with the CMB 
temperature power spectrum overplotted.  (Amplitudes of the colored lines 
are arbitrary.)  
}
\label{fig:binsandell}
\end{figure}

Figure~\ref{fig:binsandell} shows the bins in log scale factor (20 bins 
per decade) and the conversion to multipole space (overlaid with the 
CMB temperature power spectrum) by $\ell\approx \eta_0/\eta(a)$, which 
approximately relates a given wavenumber $k\approx \ell/\eta_0$ to the time 
it entered the horizon.  Note 
that a uniform binning in $\log a$, which is the expected characteristic 
scale for physical variations in the expansion, is not uniform in multipole 
space.

%%%%%%%%%%%%%%%%%%%%%%%%%%%%%%%%

To place constraints on allowed deviations $\delta(a)$ we carry out a 
Fisher matrix calculation.  The Fisher matrix elements are given by 
\begin{equation}
F_{ij} = \sum_{l} \sum_{X,Y} \frac{\partial{C_{Xl}}}{\partial{p_{i}}} (COV_{l})^{-1}_{XY} \frac{\partial{C_{Yl}}}{\partial{p_{j}}}\label{eq:def_FIM}
\end{equation} 
where $X,Y$ is any combination of the CMB temperature power spectrum (T), 
E-mode polarization power spectrum (E), and temperature-polarization 
cross power spectrum (TE).  The covariance matrix $COV$ is given by the 
measurement uncertainties of the CMB observations; we adopt the 
characteristics of the Planck satellite experiment \cite{planck}.  
The parameter set $\{p_i\}$ includes the usual cosmological parameters -- 
the physical baryon density $\Omega_b h^2$, physical cold dark matter 
density $\Omega_c h^2$, total present matter density $\Omega_m$ (the 
present Hubble constant $h$ is a derived quantity), primordial 
scalar perturbation power law index $n_s$, optical depth $\tau$, and 
present amplitude of mass fluctuations $\sigma_8$ -- and the $N_{\rm bin}$ 
expansion variation parameters $\delta(a_i)$.  The uncertainties on each 
$\delta(a_i)$ are given by the (square root of the) respective diagonal 
element of the inverse of the Fisher matrix. 

\begin{figure*}[htbp!]
\includegraphics[width=\textwidth]{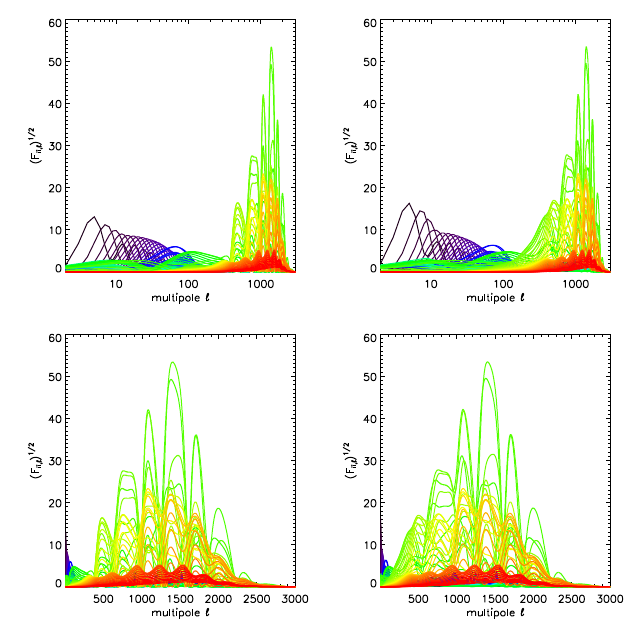}
\caption{The fractional Fisher sensitivities, $\sqrt{F_{ii,\ell}}$, to the 
expansion variations 
$\delta(a_i)$ in each bin, color coded as in Fig.~\ref{fig:binsandell}, 
are plotted vs multipole.  The top panels use log scale in multipole, the 
bottom panels a linear scale to highlight different regions. The left 
panels are for the temperature power spectrum only, while the right panels 
use the variance weighted sum of the T, E, and TE power spectra entering the 
Fisher information matrix.  Polarization information fills in the sensitivity 
gaps due to the acoustic troughs. 
}
\label{fig:binCl}
\end{figure*}

Figure~\ref{fig:binCl} shows the sensitivity $\sqrt{F_{ii,\ell}}$ of the 
weighted combination of CMB power spectra (T, E, TE) 
to the expansion deviations in each redshift bin for each multipole.  
Sensitivity 
peaks around the acoustic peaks and is reduced at low multipoles due to 
cosmic variance and at high multipoles due to the finite resolution from 
the instrument beam size.  Since the polarization power spectrum is 
out of phase with the temperature power spectrum, dips in the temperature 
sensitivity are filled in by polarization information.

%%%%%%%%%%%%%%%%%%%%%%%%%%%%%%%%                       

\begin{figure*}[htbp!]
\includegraphics[width=\textwidth]{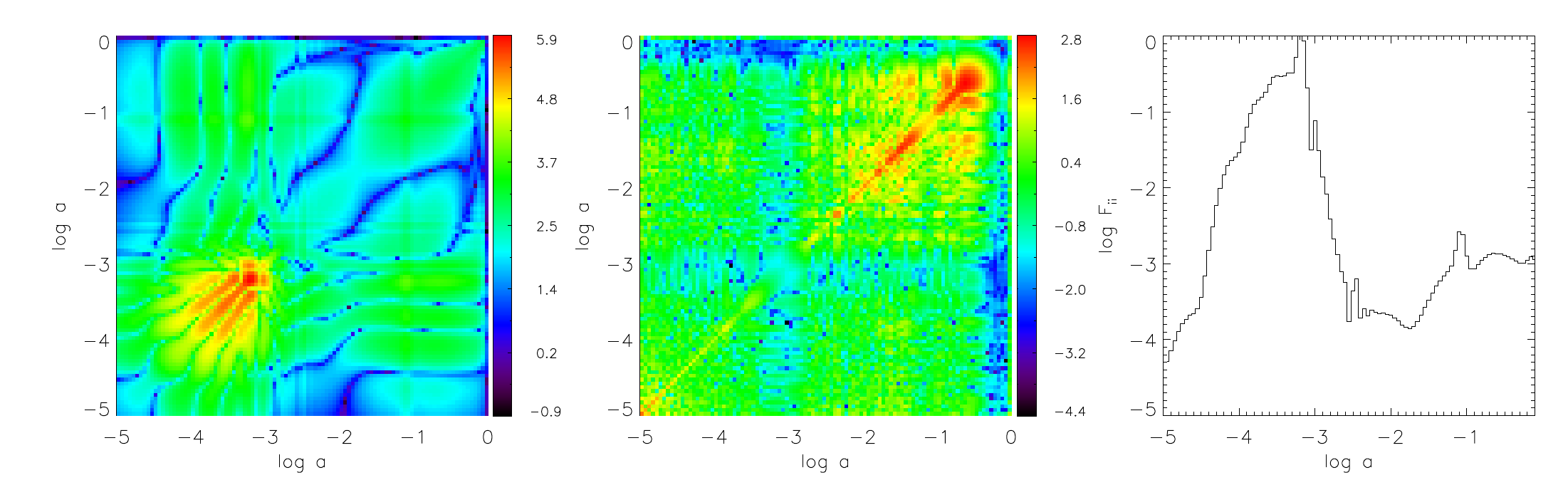}
\caption{The Fisher information submatrix (left) corresponding to the 
expansion variation parameters $\delta(a_i)$ is plotted with redder 
shades representing larger absolute values (more information).  
The color bar gives the log of the absolute value of elements.  
The covariance matrix (middle), marginalized over other cosmological 
parameters, follows the same color scheme, so the best determined 
parameters (smallest errors) are bluest.  Diagonal elements of the 
Fisher matrix (right), as a fraction of the largest diagonal element, 
quantify for which redshifts the CMB data is most sensitive to the 
expansion history.  The bump at $\log a\approx-1.1$ reflects reionization. 
}
\label{fig:binFisher}
\end{figure*}

Figure~\ref{fig:binFisher} shows the actual Fisher and covariance 
submatrices corresponding to the expansion bin parameters (marginalized 
over other parameters in the case of the covariance matrix).  First, we 
notice the maximum of the information content is near decoupling 
($\log a\approx -3$), as expected.  Earlier times, $\log a<-4$, map to 
multipoles on the damping tail and so have less leverage, while recent 
times, $\log a>-3$, are on the Sachs-Wolfe plateau and again have limited 
information.  The Fisher matrix is not diagonal because expansion deviations 
affect all later times, e.g.\ perturbation evolution once the wavemode is 
within horizon and integral quantities such as the sound horizon.  This 
will be one of the motivating factors for carrying out principal component 
analysis (PCA) in Sec.~\ref{sec:pca}.  The covariance matrix (inverse of 
the Fisher matrix), however, has a more diagonal structure, and so bins 
can be a useful parameter set if carefully chosen (see Sec.~\ref{sec:concl}).

%%%%%%%%%%%%%%%%%%%%%%%%%%%%%%%%                 

\begin{figure}[htbp!]
\includegraphics[width=\columnwidth]{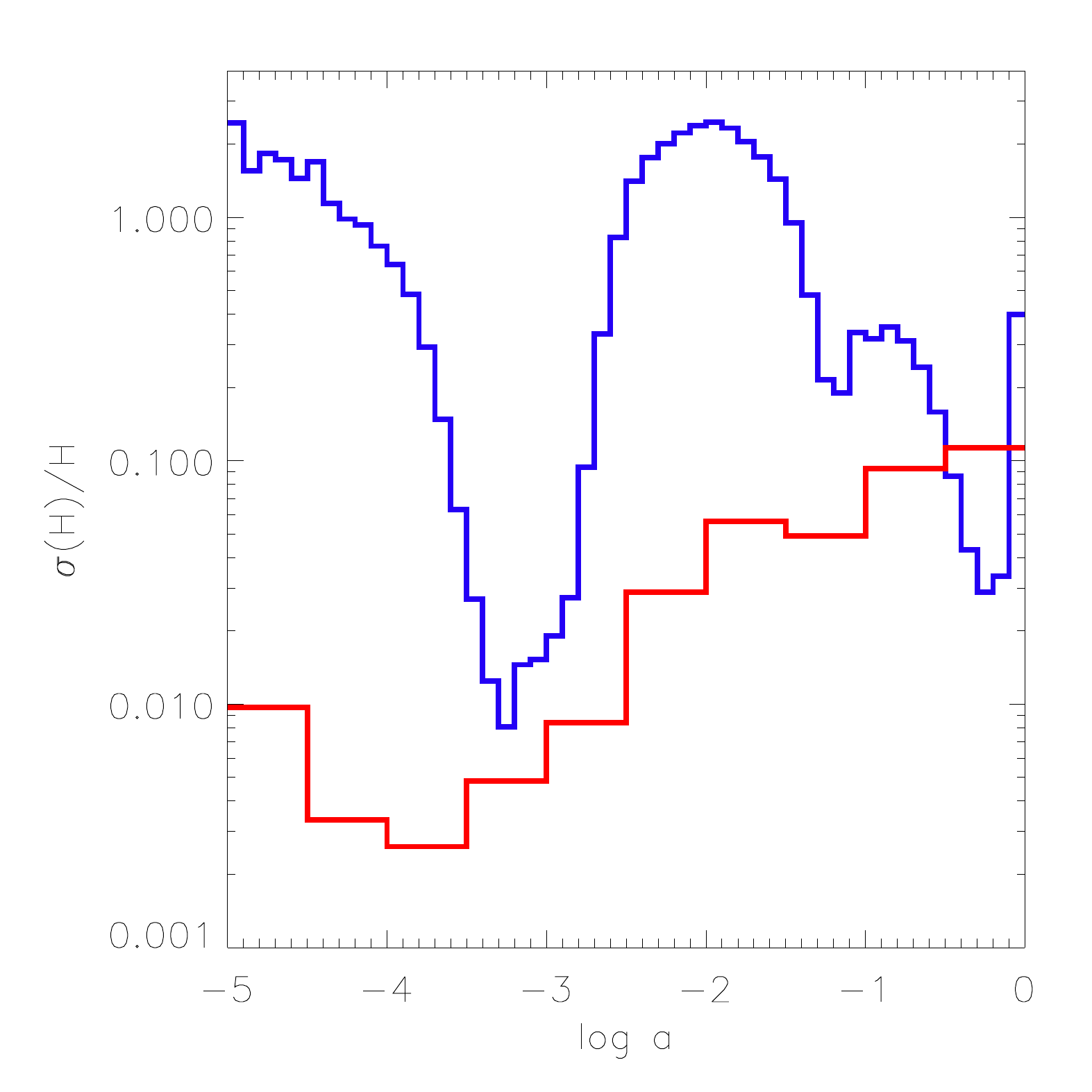}
\caption{The fractional precision with which the expansion history can 
be determined by projected Planck CMB data is plotted vs scale factor, 
for two different bandwidths.  The top (bottom) curve is for 10 (2) bins per 
decade in scale factor.  Subpercent precision can be achieved around 
decoupling but large swaths of the cosmic history will remain unknown. 
}
\label{fig:binCov}
\end{figure}

Most importantly, Figure~\ref{fig:binCov} shows the constraints on the 
expansion history 
\be 
\frac{\sigma(H)}{H_{\rm fid}}=\frac{\sigma(\delta)}{2\sqrt{1+\delta}} \ , 
\ee 
i.e.\ the fractional uncertainty on $H(a)$ due to deviations $\delta$, 
marginalized over the other cosmological parameters.  
This is the ``textbook'' plot, showing the state of our knowledge of the 
early expansion history of the universe when given CMB data of Planck 
sensitivity.  
The constraints depend on what bandwidth we wish to constrain the 
expansion history: the top curve shows 10 bins per decade, the bottom 
curve 2 bins per decade.  One can trade off 
sensitivity to fine features vs overall level of constraints.
With 10 
bins per decade one can achieve percent level constraints on $H^2(a)$ near 
decoupling, while with 2 bins per decade one obtains subpercent constraints 
over more than two decades in scale factor.  The relation between 10 bins and 
2 bins is not simply a $\sqrt{5}$ scaling due to correlations between bins 
(the offdiagonal elements of the covariance matrix), and the lowest redshift 
bin is particularly affected by covariance with the other cosmological 
parameters.

%%%%%%%%%%%%%%%%%%%%%%%%%%%%%%%%                    

%%%%%%%%%%%%%%%%%%%%%%%%%%%%%%%%%%%%%%%%%%%%%%%%%%%%%%%%% 
\section{Principal Components of Expansion} \label{sec:pca} 

Expansion deviations at some redshifts may have substantially the same 
consequence for the observables as deviations at some other redshift, or 
deviations may be correlated in such a way that only the difference 
between them is important.  This leads to the idea of compressing the 
100 bins between $\log a=[-5,0]$, or at least the information contained 
in them, into fewer variables.  One might for example speculate that the 
major effect of expansion deviations for $a>10^{-3}$ comes from shifting 
the distance to CMB last scattering, and whether the variation occurs at 
$a=0.01$ or $a=0.1$ is less crucial. 

Principal component analysis (PCA) can provide an efficient way to compress 
the influence on the observables.  For some uses probing dark energy and 
the CMB, see for example \cite{huoka,hutstar,leach,reion,tobin,dvor}  
(and dark matter and the CMB in \cite{fink}).  By diagonalizing the 
Fisher matrix we can find its eigenvectors that best summarize the 
sensitivity of the observations to the expansion deviations.  We can then 
transform the bin basis to an orthogonal eigenmode basis of principal 
components (PCs), writing 
\be 
\delta(a)=\sum m_i\,e_i(a) \ , 
\ee 
where $m_i$ is the amplitude of mode $i$ and $e_i(a)$ is the eigenvector.  
Since the modes are orthogonal, the errors $\sigma_i\equiv\sigma(m_i)$ 
on the amplitudes 
are uncorrelated.  Using the entire set of bins or the entire set of PCs is 
equivalent, but using only a few PCs with the highest eigenvalues (smallest 
$\sigma_i$) in general allows one to approximate the full set more accurately 
than the same number of bins.  That is, the information can be compressed. 

\begin{figure}[htbp!]
\includegraphics[width=\columnwidth]{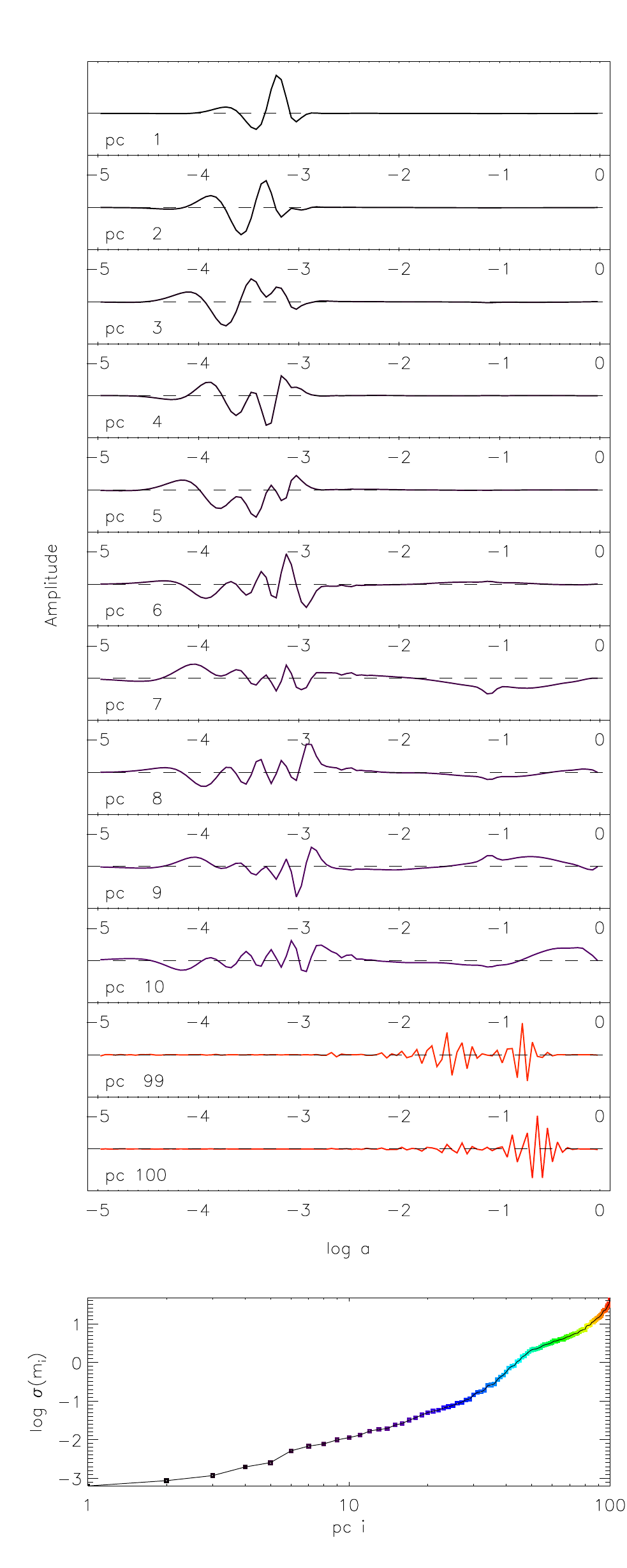}
\caption{Principal components of the CMB observational sensitivity to 
expansion history are plotted vs $\log a$, for the first 10 and last 2 
modes.  The bottom plot shows the $\sigma_i$ for all modes. 
}
\label{fig:pcs}
\end{figure}

Figure~\ref{fig:pcs} illustrates some of the PCs for the CMB Fisher matrix, 
ordered from highest to lowest eigenvalues (best to worst determined). 
Note that as expected most of the activity in the first PCs is 
prerecombination, associated with the acoustic peaks.  In modes 7-10, there 
is some low redshift action coming primarily from the integrated distance 
to last scattering and the reionization epoch.  Higher PC modes tend to be 
more oscillatory (essentially high derivatives of the expansion behavior) 
and localized.

%%%%%%%%%%%%%%%%%%%%%%%%%%%%%%%%                

By taking the cumulative sum of the eigenvalues, we find that the first 
7 PCs contain 99\% of the variance. 
That is, $\sum_1^7 \sigma_i^{-2}/\sum_1^{100} \sigma_i^{-2}=0.99$.  This 
means that the great majority of expansion behaviors, as far as their 
observational detectability is concerned, can be described with just 
7 parameters $m_1,\dots m_7$.  

It is convenient to normalize the PCs such that 
$\sum_n e_i(a_n) e_j(a_n)=\delta_{ij}$, where $\delta_{ij}$ here is the 
Kronecker delta, 
and then the mode coefficients $m_i$ give the amplitudes for a given model 
of expansion history deviation, 
\be 
m_i = \sum_n \delta(a_n) e_i(a_n) \ , \label{eq:mi} 
\ee 
where $a_n$ denotes the bin centers.  For many narrow bins the sum can 
be converted to an integral.  The amplitudes $m_i$ only have meaning 
when discussing a specific model; note the canonical model $\Lambda$CDM 
has all $m_i=0$.  Since the mode amplitude $m_i$ has no a priori magnitude, 
we do not know whether its uncertainty $\sigma_i=0.01$, say, is a small or 
large number.  
We can compare $\sigma_i$'s to each other (but if the $m_i$ are 0 this is 
irrelevant), or for a specific model we can form a signal to noise ratio for 
a mode, 
\be 
(S/N)_i=m_i/\sigma_i \ . \label{eq:sn} 
\ee 
If $(S/N)_i\ll1$, then it does not matter if $\sigma_i$ looks small, the 
mode cannot be well measured.  Thus caution must be used in interpreting 
PCA (for more details see \cite{tobin,beingpc}). 

Although PCA certainly compresses the information content into fewer 
parameters, the number of PCs required to describe arbitrary expansion 
histories is still large for the purposes of, say, Monte Carlo simulations.  
We can therefore use PCA instead as a guide in two ways: to examine the 
ability to discriminate among different classes of models for the expansion 
deviation, 
and to indicate which regions in $a$ show the most sensitivity to deviations 
and hence which of the original bins are most useful.  These are treated 
respectively in Sec.~\ref{sec:edemodels} and Sec.~\ref{sec:concl}.

%%%%%%%%%%%%%%%%%%%%%%%%%%%%%%%%%%%%%%%%%%%%%%%%%%%%%%%%% 
\section{Comparing Early Dark Energy Models} \label{sec:edemodels} 

While we wish to concentrate, as much as possible, on a model independent 
approach to constraining the expansion history, it is useful to make contact 
with various classes of models to make sure that important behaviors are 
captured.  In broad strokes, one can consider cases where the expansion 
deviations decline at times earlier than recombination, increase at earlier 
times, or remain constant.  This can be translated into early dark energy 
models that contribute a lower fraction of the total energy density in 
radiation vs matter domination, a higher fraction, or a constant fraction. 
This tilt of the expansion rate can be an important discriminant, somewhat 
analogous to the tilt of the primordial power spectrum for inflation.  

Early dark energy models have been proposed with each of these behaviors. 
Examples of the three classes respectively are 1) Barotropic dark energy 
with sound speed $c_s^2=0$ \cite{linscher}, sometimes called aether models, 
2) Barotropic dark energy with sound speed $c_s^2=1/3$ \cite{linscher}, 
sometimes called dark radiation, and 3) Scaling dark energy such as the 
commonly used Doran-Robbers model with $c_s^2=1$ \cite{dorrob}.  

Barotropic models have $w(a)$, $\rho_{de}(a)$, and hence $\delta(a)$, all 
determined by $c_s^2(a)$.  The dynamics is given by \cite{linscher} 
\be 
w'=-3(1+w)(c_s^2-w) \ , 
\ee 
with solution for constant $c_s$ of 
\be 
w(a)=\frac{c_s^2 Ba^{-3(1+c_s^2)}-1}{Ba^{-3(1+c_s^2)}+1} \ , 
\ee 
where $B=(1+w_0)/(c_s^2-w_0)$.  For the $c_s^2=0$ aether model, 
\bea 
\rho_{ae}(a)&=&\rho_\infty+\Omega_e\rho_{m,0}a^{-3} \\ 
\delta_{ae}(a)&=&\Omega_e\,\frac{\Omega_m(a^{-3}-1)}{\Omega_m (a^{-3}- 
1)+ \Omega_r a^{-4} +1} \\ 
\rho_\infty/\rho_{crit,0}&=&1-\om(1+\Oe)=(-w_0)(1-\om) \\ 
w_0&=&-1+\frac{\Omega_e\om}{1-\om} \ , 
\eea 
where during matter domination the dark energy contributed a constant 
fractional density $\Omega_e$, but this declines at earlier times 
as radiation becomes important.  

For the $c_s^2=1/3$ dark radiation model, 
\bea 
\rho_{dr}(a)&=&\rho_\infty+\Omega_e \rho_{r,0}a^{-4} \\ 
\delta_{dr}(a)&=&\Omega_e\,\frac{\Omega_r(a^{-4}-1)}{\Omega_r(a^{-4}-1)+\om 
(a^{-3}-1)+1} \\ 
\rho_\infty/\rho_{crit,0}&\approx&1-\om \\ 
w_0&\approx& -1 \ , 
\eea 
where during radiation domination the dark energy contributed a constant 
fractional density $\Omega_e$, but this declines at later times
as matter becomes important.  

The most commonly used early dark energy is the Doran-Robbers form 
\cite{dorrob}, 
\be 
\Omega_{de}(a)=\frac{1-\om-\Oe(1-a^{-3w_0})}{1-\om(1-a^{3w_0})} 
+\Oe(1-a^{-3w_0}) \ , 
\ee 
where during both matter and radiation domination the dark energy 
contributed a constant fractional density $\Oe$.  The sound speed is 
conventionally taken to be $c_s^2=1$.  

Thus the three models we consider have expansion history deviations 
with complementary behaviors: rising, falling, and constant.  This 
range should give a good indication of how PCA can characterize the 
expansion, while also having physical motivations.  Of the physics origins 
mentioned in the Introduction, some string theories give Doran-Robbers 
behavior, some noncanonical kinetic fields give aether behavior, and 
some higher dimension theories give dark radiation.  

Once we calculate 
the $m_i$ for a model through Eq.~(\ref{eq:mi}), we have a better indication 
of the importance of a PC mode through the signal to noise criterion of 
Eq.~(\ref{eq:sn}) -- recall that the $\sigma_i$ alone say little about 
whether the mode is relevant.  To choose the number of modes to keep in 
describing a model, we can simply impose a $S/N$ cutoff, or ask that the 
cumulative $S/N$ of the modes kept be above some fraction, e.g.\ 95\% 
of the total $S/N$ from all modes. 

Another indicator is the statistical risk, or mean squared error.  This 
takes into account that more modes decrease the bias from the true model, 
but increase the accumulated variance in $\delta$.  The risk $R_N$ is 
\bea 
R_N&=&\sqrt{b_N^2+\sigma_N^2} \\ 
b_N^2&=&\sum_n [\delta^{\rm model}(a_n)-\delta^{N\,{\rm PCs}}(a_n)]^2 \\ 
\sigma_N^2&=&\sum_n \sum_{i=1}^N \sigma_i^2 e_i^2(a_n) \ . 
\eea 
One could choose to keep that number $N$ PCs where $R_N$ is minimized.  

\begin{figure*}[htbp!]
\includegraphics[width=\textwidth]{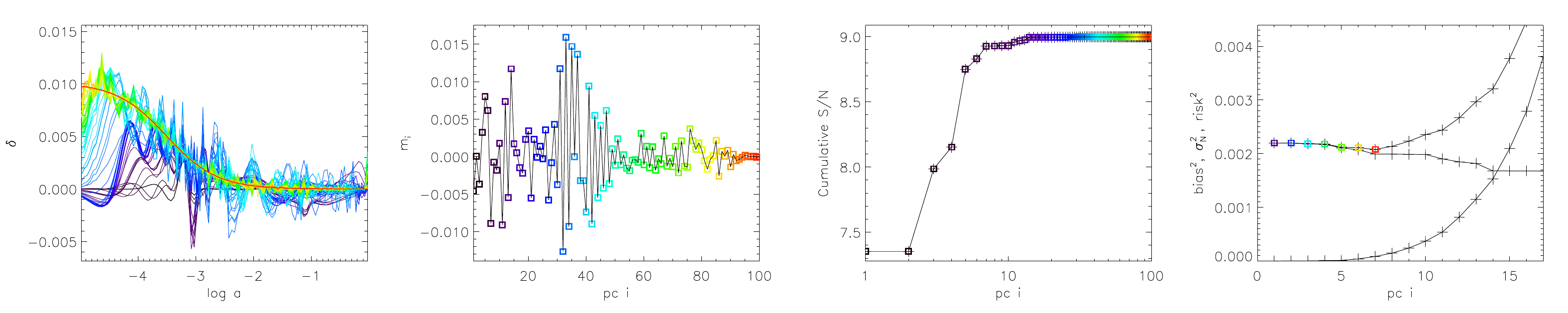}
\includegraphics[width=\textwidth]{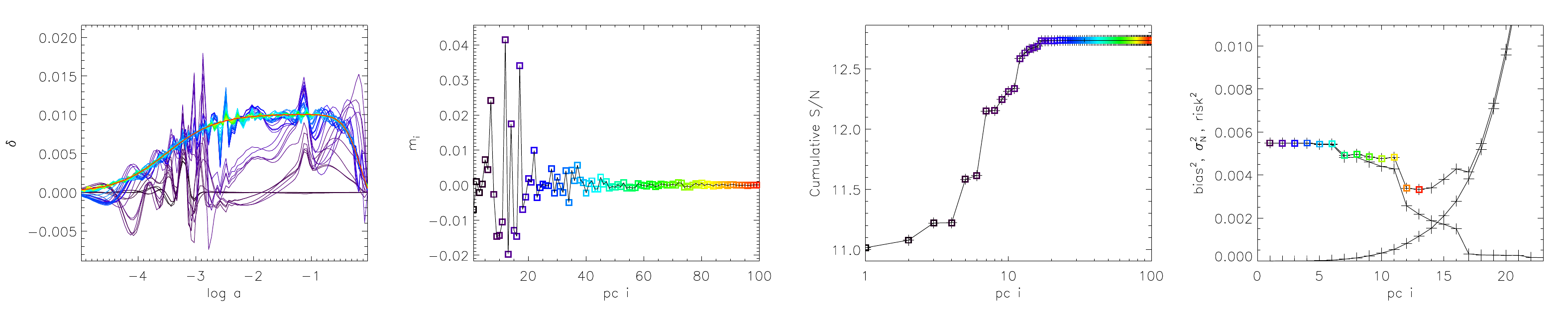}
\includegraphics[width=\textwidth]{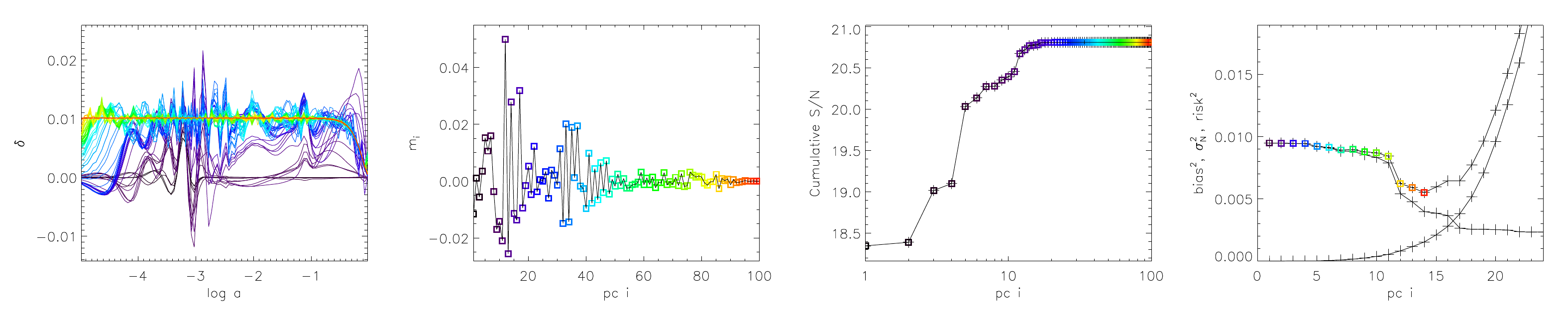}
\caption{Three models of early dark energy are analyzed -- dark radiation 
(top row), aether (middle row), and Doran-Robbers (bottom row) -- 
with different recombination era behaviors.  The leftmost column shows 
$\delta(a)$ built up out of PCs, with the thick red line giving the exact 
model.  The second column gives the PC amplitudes $m_i$ and the third 
column shows the cumulative $S/N$, summing Eq.~(\ref{eq:sn}) over the 
first $i$ PCs in quadrature.  The rightmost column shows the bias squared 
(falling curve), variance (rising curve), and risk squared (top curve) when 
including the first $i$ 
PCs.  One might choose to keep those PCs that either give the largest 
jumps in $S/N$, or all those up to the minimum in the risk curve. 
}
\label{fig:modelpca}
\end{figure*}

Figure~\ref{fig:modelpca} shows for each of the three models the 
approximation to $\delta(a)$ as more PCs are added, the values $m_i$, 
and the cumulative $S/N$ and risk as a function of number of PCs used. 
In general we find the risk requires more PCs than the $S/N$ criterion; 
this makes sense since $S/N$ concentrates on those PCs fitting the 
observable (CMB power spectra) while risk attempts to fit the unobservable 
expansion deviation.  Another drawback to risk is that it does not scale 
with the amplitude of the modes, i.e.\ while $S/N$ increases linearly with 
$\Oe$, the bias term in the risk scales but the variance 
does not, so the risk is weighted unevenly depending on deviation 
amplitude. 

For the dark radiation, 
aether, and Doran-Robbers models, respectively, we should keep the first 
6, 7, and 5 PCs according to $S/N$, and 10, 14, and 15 PCs according to 
risk.  However, we note that this is if we keep the PCs in order according 
to $\sigma_i$.  If we choose the highest $S/N$ modes individually, we 
only require 4, 3, and 3 modes to attain 95\% of the full $S/N$ (but this 
requires knowing the correct model, or assuming a given set of models).

%%%%%%%%%%%%%%%%%%%%%%%%%%%%%%%%                           

\begin{figure}[htbp!]
\includegraphics[width=\columnwidth]{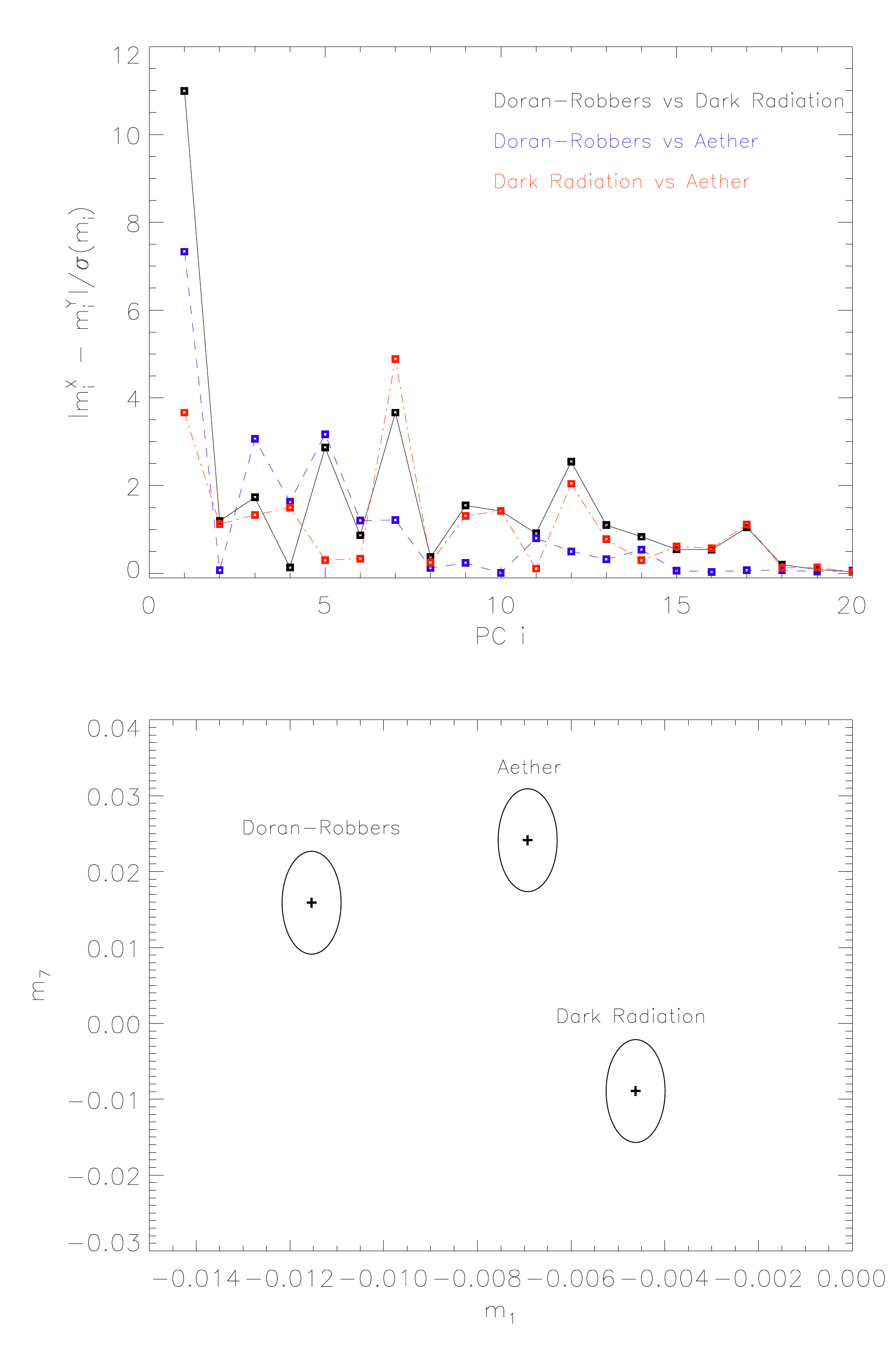}
\caption{(Top) Differences between each PC amplitude $m_i$ are shown 
for pairs $(X,Y)$ of early dark energy models, with solid curves comparing 
Doran-Robbers to dark radiation, dashed Doran-Robbers to aether, and 
dot-dashed dark radiation to aether.  The highest peaks indicate the 
modes with strong discriminating power.  (Bottom) Amplitudes of 
modes 1 and 7 are plotted for the three models, with the +'s indicating 
the values $m_i$ and the ellipses showing the uncertainties $\sigma_i$.  
These two modes can clearly distinguish between each of the three models.
}
\label{fig:modelmimj}
\end{figure}

Let us examine these models in more detail.  Figure~\ref{fig:modelmimj} 
shows how these models are well separated in eigenmode coefficient space. 
Considerable discriminating power occurs using modes 1 and 7, for example, 
with the separations between the three models many times the uncertainties 
$\sigma_i$.  From the shape of the modes in Fig.~\ref{fig:pcs}, we can 
see that mode 1 is roughly measuring the amplitude of the expansion deviation 
at recombination, and whether it is increasing or decreasing (thus 
distinguishing all three models), and mode 7 is sensitive to more recent 
deviations such as occurring in Doran-Robber and aether, but not dark 
radiation, cases.  Thus these two modes together are adept at distinguishing 
between the rising, falling, and constant deviation classes of expansion 
history, and early vs late deviations.

%%%%%%%%%%%%%%%%%%%%%%%%%%%%%%%%                           

\begin{figure*}[htbp!]
\includegraphics[width=\textwidth]{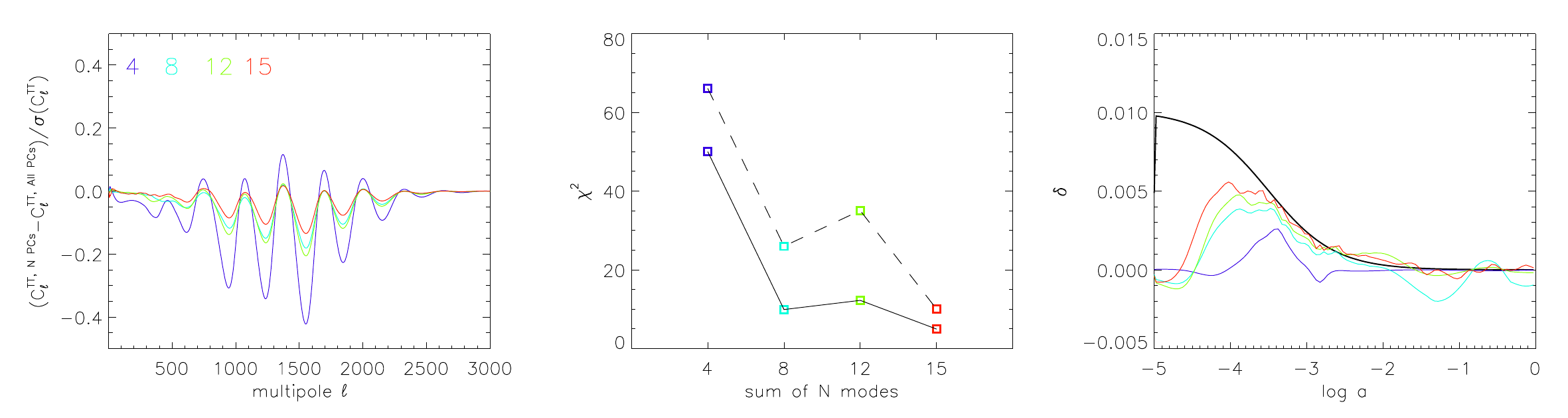}
\includegraphics[width=\textwidth]{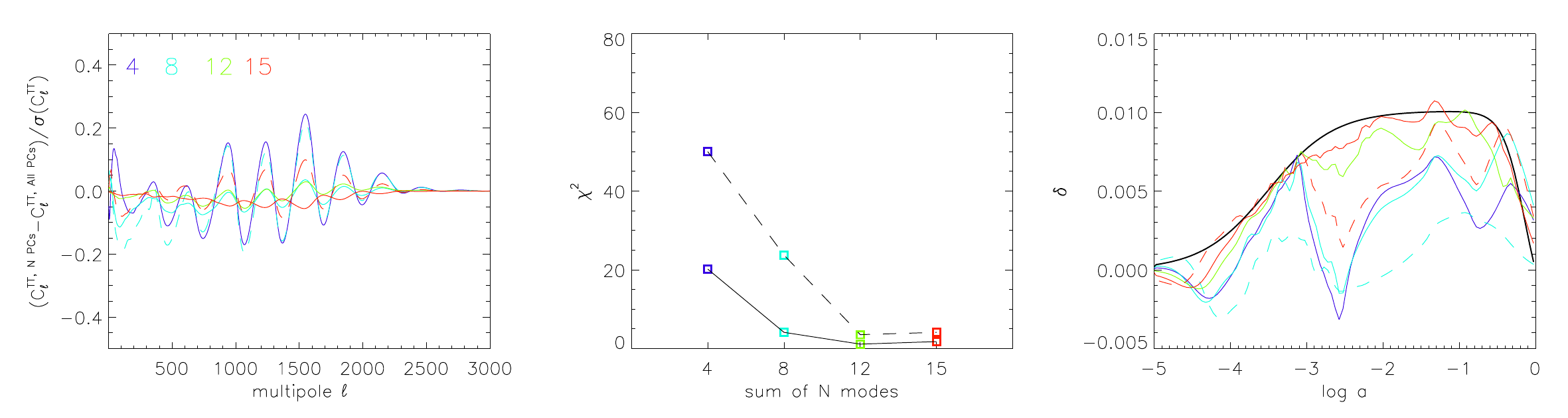}
\includegraphics[width=\textwidth]{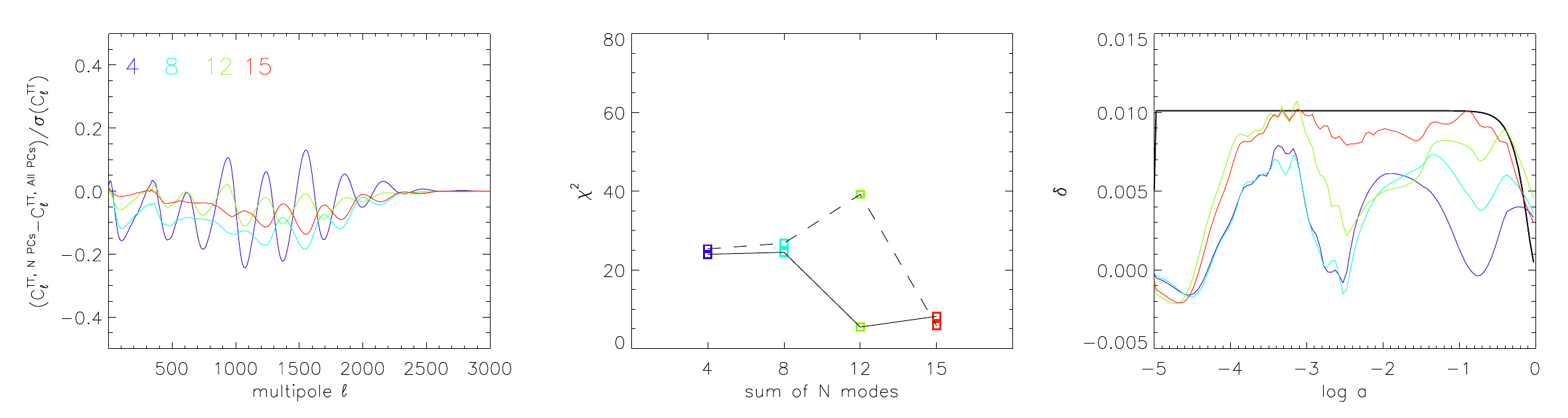}
\caption{PCA of three models of early dark energy -- dark radiation 
(top row), aether (middle row), and Doran-Robbers (bottom row) -- is 
compared to the exact models in terms of the observable CMB power spectrum.  
The left column shows the deviation in the temperature power spectrum for 
the sum of the first 4, 8, 12, 15 modes using $S/N$ ordering of the modes 
(see Fig.~\ref{fig:modelpca}).  The $\chi^2$ of the deviations summing over 
multipoles is in the middle panels, with the dashed curve using ordering 
by $\sigma_i$ instead.  Reconstruction of the theoretical $\delta(a)$ 
appears in the right panels; note how only certain epochs need be fit 
well to reproduce the observable CMB.  For the aether model (middle row) 
dashed curves show also the $\sigma_i$ ordering results for the $N=8$, 
15 cases, which do much worse than $S/N$ ordering. 
}
\label{fig:clpcs}
\end{figure*}

We emphasize that the PCs are really telling us about fits to the 
observable power spectra and not reconstruction of the expansion deviation 
in a fine grained sense.  While Fig.~\ref{fig:modelpca} shows that 
$\sim50$ PCs are needed to model $\delta(a)$ well for these example, 
Fig.~\ref{fig:clpcs} demonstrates that only $\sim10$ PCs are needed to 
give accurate 
agreement in $C_\ell$.  Figure~\ref{fig:clpcs} shows the residual of 
the CMB temperature power spectrum from the sum of the first $N$ PCs 
(in $S/N$ ordering, and also shown in $\sigma_i$ ordering for the aether 
model) relative to the true model for the three models, as well as the 
$\chi^2$ summed over multipoles, and the reconstructed expansion history.  
Although the reconstructed expansion history may only agree over certain 
ranges of $a$, this can still give excellent agreement for the observables 
as the power spectra are not equally sensitive to all scale factors.

%%%%%%%%%%%%%%%%%%%%%%%%%%%%%%%%                     

The numbers of PCs required to ensure an accurate estimation of the 
observables, say $\chi^2<10$ (summed over 3000 multipoles), is generally 
greater than 10, rendering cumbersome a straight application of mode 
amplitudes as parameters in a Monte Carlo simulation.  (And recall that 
for a model independent analysis we do not know a priori the $S/N$ 
ordering, so one would need to include many more PCs using $\sigma_i$ 
ordering.)  Moreover, PCs per se are not always easy to interpret in terms 
of physical effects.  Therefore it is both clearer and more efficient to 
use the PCA instead to guide an effective, low order binning basis.

%%%%%%%%%%%%%%%%%%%%%%%%%%%%%%%%%%%%%%%%%%%%%%%%%%%%%%%%% 
\section{Conclusions} \label{sec:concl} 

Our knowledge of the expansion history of our universe, even at the 
level of degree of matter domination or radiation domination at early 
epochs, is remarkably imprecise.  Cosmic microwave background radiation 
measurements from ACT, Planck, and SPT (and later ACTpol and SPTpol) 
will shed light on the times around recombination and reionization.  
We quantify the model independent state of our knowledge through a 
combination of redshift bin and principal component analysis, 
finding that subpercent level constraints will be 
placed by Planck over $\log a=[-2.5,-5]$ for a bandwidth of 
$\Delta\log a=0.5$. 

CMB data will address one of the key aspects of dark energy -- its 
persistence, a characteristic of many high energy physics origins -- 
and we find that several different classes of early dark energy are well 
separated in PCA space.  The limits can also be interpreted in terms of 
the number of effective relativistic species, $\neff$, such as an extra 
neutrino type, with current data mildly preferring further contributions. 
A thermal relativistic neutrino species adds 23\% to the photon energy 
density, so $\delta=0.13\,\Delta\neff$, giving tight limits on extra 
relativistic degrees of freedom from the forthcoming data.  

We explore three specific models, representing different classes for 
early dark energy, possibly corresponding to different physical origins.  
The commonly used Doran-Robbers form has a dark 
energy fraction $\Omega_e$ that is constant through the recombination epoch.  
We also investigate a dark radiation model with $\Omega_{de}(a)$ rising to 
the past and a barotropic aether model with $\Omega_{de}(a)$ falling to
the past, and find that the dominant PC mode is well able to distinguish 
between these behaviors.  Since the amplitude of that mode is greatest 
for the Doran-Robbers model, we expect that data constraints on $\Omega_e$ 
in the other classes will be weaker than in this model (such as from 
\cite{reichardt} using current CMB data), allowing for nonnegligible 
persistence of dark energy (see \cite{rdof} for further demonstrations 
of this).  
Our general approach, however, does not rely on assuming the form for the 
new component or expansion deviation. 

Figure~\ref{fig:binFisher} is in a sense the textbook picture of what 
Planck CMB data will say in a model independent manner about early universe 
expansion.  For postrecombination 
epochs this will improve with further ground based polarization measurements 
(especially of CMB lensing) and inclusion of growth of structure data. 
Understanding early expansion is in fact crucial for accurate interpretation 
of large scale structure, and feeds directly into the early time 
gravitational growth calibration parameter $g_\star$ \cite{gstar}; 
ignorance of this 
can bias cosmological parameter estimation and tests of gravity. 

Expansion history is not the whole story as the effective fluid behind the 
expansion deviations has perturbations and can have internal degrees of 
freedom.  We treat the perturbations consistently -- the dark radiation 
and barotropic aether models for 
example have sound speeds different from the speed of light.  We do not 
include viscosity, however, as the data has poor leverage on this 
\cite{cold,rdof,smithdaszahn}.  Another difficulty for model independent 
analysis is having $\delta<0$, since perturbations are difficult to treat 
when the effective density deviation passes through zero; models such as 
nonthermal neutrinos, with energy densities below the standard, could 
realize such a condition.  We will consider such cases in future work. 

Principal component analysis provides a valuable guide to the key epochs 
of sensitivity and the amount of information contributed from different 
times.  However, we emphasize and demonstrate that the raw uncertainty 
$\sigma_i$ on an eigenmode has very limited meaning; the first 15 modes 
ordered by $\sigma_i$ can give a highly inaccurate reconstruction relative 
to a smaller number of modes ordered by signal to noise.  Redshift bins 
can be more clearly interpreted.  Employing the best aspects of each can 
result in physically clear, well characterized expansion history 
constraints.

%%%%%%%%%%%%%%%%%%%%%%%%%%%%%%%%%%%%%%%%%%%%%%%%%%%%%%%%%%%%    

\acknowledgments

This work has been supported by World Class University grant 
R32-2009-000-10130-0 through the National Research Foundation, Ministry 
of Education, Science and Technology of Korea and the Director, 
Office of Science, Office of High Energy Physics, of the U.S.\ Department 
of Energy under Contract No.\ DE-AC02-05CH11231. 
The Dark Cosmology Centre is funded by the Danish National Research 
Foundation.

%%%%%%%%%%%%%%%%%%%%%%%%%%%%%%%%%%%%%%%%%%%%%%%%%%%%%%%%%%%%

\end{document}